\documentclass{article}
\usepackage{scalefnt,spconf,amsmath,graphicx,amssymb}
\usepackage{graphicx}
\usepackage{caption}
\usepackage{placeins}
\usepackage{algorithm,algpseudocode}
\usepackage{cite}

\pdfmapfile{=pdftex.map}

\title{A Parametric Non-negative Coupled Canonical Polyadic Decomposition Algorithm for Hyperspectral Super-Resolution}
\name{Xi-Yuan Liu, Xiao-Feng Gong, Lei Wang, Wei Feng, Qiu-Hua Lin\thanks{ This work is supported by National Natural Science Foundation of China
 under grants 62071082, 61871067, 62471082 and 62471084, and China Postdoctoral
 Science Foundation under grant 2020M680922.}}
\address{ School of Information and Communication Engineering, Dalian University of Technology, China\\
 E-mails: xfgong@dlut.edu.cn}
\begin{document}
%
\maketitle
\begin{abstract}
Recently, coupled tensor decomposition has been widely used in data fusion of a hyperspectral image (HSI) and a multispectral image (MSI) for hyperspectral super-resolution (HSR). However, exsiting works often ignore the inherent non-negative (NN) property of the image data, or impose the NN constraint via hard-thresholding which may interfere with the optimization procedure and cause the method to be sub-optimal. As such, we propose a novel NN coupled canonical polyadic decomposition (NN-C-CPD) algorithm, which makes use of the parametric method and nonlinear least squares (NLS) framework to impose the NN constraint into the C-CPD computation. More exactly, the NN constraint is converted into the squared relationship between the NN entries of the factor matrices and a set of latent parameters. Based on the chain rule for deriving the derivatives, the key entities such as gradient and Jacobian with regards to the latent parameters can be derived, thus the NN constraint is naturally integrated without interfering with the optimization procedure. Experimental results are provided to demonstrate the performance of the proposed NN-C-CPD algorithm in HSR applications.  

\end{abstract}
\section{Introduction}
\label{sec:intro}

Recently, coupled tensor decomposition techniques \cite{STEREO,nonCCPD,C-Tucker,TD-TUCKER-NN,C-BTD,CBTDNN2024} have attracted great attention in multiset data fusion of a hyperspectral image (HSI) and a multispectral image (MSI) to get super-resolution image (SRI) \cite{review}.

In \cite{STEREO}, an unconstrained coupled canonical polyadic decomposition (C-CPD) algorithm based on alternating least squares (ALS) was proposed for hyperspectral super-resolution (HSR). In \cite{nonCCPD}, a nonlocal C-CPD algorithm based on alternating direction method of multipliers (ADMM) method \cite{ADMM} was proposed for HSI-MSI data fusion. In \cite{C-Tucker}, HSI and MSI are approximated to the coupled Tucker (C-Tucker) model. With the degradation mechanism from SRI to HSI/MSI established as the flexible coupling \cite{FCCPD} between HSI and MSI datasets, these coupled tensor based techniques can well exploit the coupling structure between HSI and MSI datasets, thereby achieving accurate reconstruction of SRI. However, the inherent non-negative (NN) feature of HSI and MSI data is mostly ignored in these methods. 

In \cite{NNMATRIX}, an NN coupled matrix decomposition method was proposed for HSR. Adding NN constraint improves the stability of the algorithm \cite{NNRS}. However, this approach was based on the matrix model and did not take full advantage of the high-dimensional structure of the datasets. In \cite{TD-TUCKER-NN}, a C-Tucker decomposition based on NN and nonlocal 4-D tensor dictionary learning which considered the non-negativity and sparsity of the datasets fully was proposed. In \cite{C-BTD}, an NN coupled block term decomposition (C-BTD) method based on alternating projection gradient (APG) was proposed. The C-BTD model writes the HSI/MSI data into the sum of low multilinear rank terms, which is more flexible than C-CPD that uses rank-1 terms.  However, the NN constraint is imposed via hard-thresholding, which may interfere optimization procedure. In \cite{CBTDNN2024}, a C-BTD algorithm with minimum volume regularizations and incorporate the $\beta$-divergence was proposed for images fusion and the factor matrices were gotten via the multiplicative update method.

In this paper, we propose a novel NN-C-CPD algorithm by using a parametric method to impose NN constraint within the C-CPD nonlinear least squares (NN-C-CPD-NLS) framework \cite{NicNNNLS}. The NN constraint is reformulated as a squared relationship between the NN entries of the factor matrices and a set of latent parameters \cite{NLS,NicNNNLS}. By applying the chain rule, the gradient and Jacobian with respect to the latent parameters can be derived, and thus the NN constraint can be naturally integrated into the optimization procedure. Numerical experiments are conducted to illustrate the performance of the proposed algorithms in comparison with several existing coupled tensor/matrix based HSI/MSI fusion methods. 

{\it{Notations}}: Scalars, vectors, matrices and tensors are denoted by italic lowercase, lowercase boldface, uppercase boldface letters, and uppercase calligraphic letters, respectively. 
We use vec($ \cdot $) to represent the vectorization operation. 
$\mathbf{Y}_{(n)}$ is denoted by the mode-$\it{n}$ matrix representation of a third-order NN tensor 
${\boldsymbol{ \mathcal {Y}}} \in {\mathbb{R}^{I \times J \times K}_{+}}$. 
We use ‘$\circ$’, ‘$\otimes$’, ‘$*$’ and ‘${ \times _n}$’ to denote vector outer product, Kronecker product, Hadamard product and mode-$\it{n}$ product, respectively. 

An NN CPD expresses ${\boldsymbol{ \mathcal {Y}}} \in {\mathbb{R}^{I \times J \times K}}$ as the sum of minimal number of rank-1 terms where 
$\mathbf{A} = [{\mathbf{a}_1}, \ldots ,{\mathbf{a}_R}],\mathbf{B} = [{\mathbf{b}_1}, \ldots ,{\mathbf{b}_R}],\mathbf{C} = [{\mathbf{c}_1}, \ldots ,{\mathbf{c}_R}],$  and $\it{R}$ is defined as the NN rank of ${\boldsymbol{ \mathcal {Y}}}$.
\[{\boldsymbol{ \mathcal {Y}}} \buildrel \Delta \over = [\kern-0.15em[ \mathbf{A}, \mathbf{B}, \mathbf{C} ]\kern-0.15em]_R  = \sum\limits_{r = 1}^R {{ \mathbf{a}_r} \circ { \mathbf{b}_r} \circ { \mathbf{c}_r}}  \in {\mathbb{R}^{I \times J \times K}},\]


\section{PROBLEM FORMULATION}
\label{sec:format}


We denote the target SRI tensor by $\boldsymbol{ \mathcal {Y}}_{S} \in \mathbb{R}^{I \times J \times K}_{+}$ which approximately admits an NN CPD model \cite{compress} as :
%
\begin{equation}\label{eq:cpd}
    \boldsymbol{ \mathcal {Y}}_{S} \approx [\![ \mathbf{A}, \mathbf{B}, \mathbf{C} ]\!]_R,
\end{equation}
where $\mathbf{A} \in \mathbb{R}^{I \times R}_{+}$ and $\mathbf{B} \in \mathbb{R}^{J \times R}_{+}$ are factor matrices in spatial dimensions, $\mathbf{C} \in \mathbb{R}^{K \times R}_{+}$ is the factor matrix in the spectral dimension.

HSI and MSI are degraded versions of SRI, with high spectral resolution but low spatial resolution, and low spectral resolution but high spatial resolution, respectively. We denote the HSI and MSI datasets as $\boldsymbol{ \mathcal {Y}}_{H} \in \mathbb{R}^{I_H \times J_H \times K}_{+}$ and $\boldsymbol{ \mathcal {Y}}_{M} \in \mathbb{R}^{I \times J \times K_M}_{+}$, respectively, where $I_H < I$, $J_H < J$, $K_M < K$. The degradation from SRI to HSI and MSI can be modeled as follows:
\begin{equation}\label{degra}
    \begin{aligned}
    &\boldsymbol{ \mathcal {Y}}_{H}(:,:,k) = \mathbf{P}_1 \boldsymbol{ \mathcal {Y}}_{S}(:,:,k) \mathbf{P}_2^T,  k\!=\!1, \ldots, K,\\
    {\boldsymbol{ \mathcal {Y}}_M}(&i,j,:) \!= \!{\mathbf{P}_M}{\boldsymbol{ \mathcal {Y}}_S}(i,j,:),i\!=\!1, \ldots ,I,j\!=\!1, \ldots ,J,
    \end{aligned}
\end{equation}
where $\mathbf{P}_1\in \mathbb{R}^{I_H \times I}_{+}$ and $\mathbf{P}_2\in \mathbb{R}^{J_H \times J}_{+}$ are two matrices that perform spatial blurring plus downsampling along the horizontal dimension and vertical dimension, respectively. $\mathbf{P}_M\in \mathbb{R}^{K_M \times K}_{+}$ is a band-selection and averaging matrix in the spectral dimension. We refer to \cite{STEREO} for details.

From \eqref{degra} we know $\boldsymbol{ \mathcal {Y}}_{H}$ and $\boldsymbol{ \mathcal {Y}}_{M}$ can be written as:
\begin{equation}\label{degra2}
    \begin{aligned}
    \boldsymbol{ \mathcal {Y}}_{H}& = \boldsymbol{ \mathcal {Y}}_{S} \times_1 \mathbf{P}_1 \times_2 \mathbf{P}_2,\\
    &{\boldsymbol{ \mathcal {Y}}_M} = {\boldsymbol{ \mathcal {Y}}_S} \times _3 {\mathbf{P}_M}.
    \end{aligned}
\end{equation}

\begin{figure}[h]
    \centering
    \includegraphics[width=0.95\columnwidth]{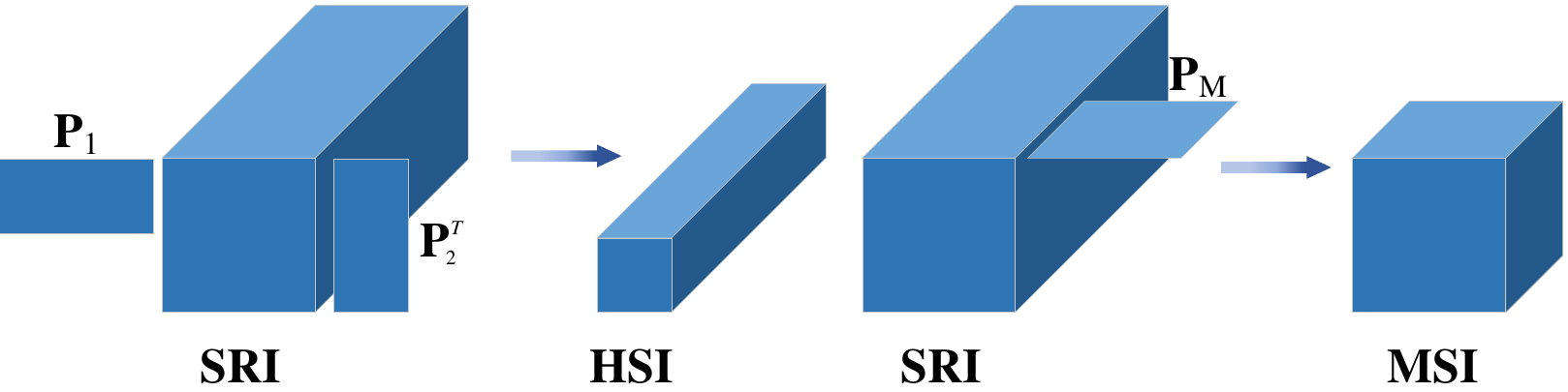}
    \captionsetup{aboveskip=0pt, belowskip=0pt} 
    \vspace{3mm} 
    \caption{The degradation of SRI}
    \label{fig:dSRI}
    \vspace{-1mm}
\end{figure}

An illustration of spatial and spectral degradation is shown in Fig. \ref{fig:dSRI}. 
As such, according to \eqref{degra}-\eqref{degra2}, the HSI and MSI datasets together admit an NN C-CPD with flexible coupling \cite{FCCPD} as follows:
\begin{equation}\label{6}
    \begin{aligned}
    {\boldsymbol{ \mathcal {Y}}_H} =  [\kern-0.15em[ {\mathbf{P}_1}\mathbf{A},{\mathbf{P}_2}\mathbf{B},\mathbf{C} ]\kern-0.15em]_R ,\\
    {\boldsymbol{ \mathcal {Y}}_M} =  [\kern-0.15em[ \mathbf{A},\mathbf{B},{\mathbf{P}_M}\mathbf{C} ]\kern-0.15em]_R,
    \end{aligned}
\end{equation}
where $\mathbf{P}_1\in \mathbb{R}^{I_H \times I}_{+}$, $\mathbf{P}_2\in \mathbb{R}^{J_H \times J}_{+}$ and $\mathbf{P}_M\in \mathbb{R}^{K_M \times K}_{+}$ are known matrices. 

\section{PROPOSED ALGORITHM}
\label{sec:pagestyle}

The NN C-CPD model in \eqref{6} can be computed by solving the following least squares optimization problem subject to NN constraint:
\begin{equation}\label{666666666}
\begin{array}{l}
\mathop {\mathrm{min}}\limits_{\mathbf{A},\mathbf{B},\mathbf{C}} {f} = {f_1} + {f_2},\\
\mathrm{s.t.}, \mathbf{A} \geq \mathbf{0},\mathbf{B} \geq \mathbf{0},\mathbf{C} \geq \mathbf{0},
\end{array}
\end{equation}
where ${f_1} = \parallel {\boldsymbol{ \mathcal {Y}}_H} -  [\kern-0.15em[ {\mathbf{P}_1}\mathbf{A},{\mathbf{P}_2}\mathbf{B},\mathbf{C} ]\kern-0.15em]_R \!\parallel _F^2,\;{f_2} = \;\parallel \!{\boldsymbol{ \mathcal {Y}}_M} -  [\kern-0.15em[ \mathbf{A},\mathbf{B},{\mathbf{P}_M}\mathbf{C} ]\kern-0.15em]_R\! \parallel _F^2$. Note that the NN constraint can be formulated as the entry-wise squared relation between $\mathbf{A}$, $\mathbf{B}$, $\mathbf{C}$ and a set of latent parameters $\mathbf{D} \in \mathbb{R}^{I \times R}$, $\mathbf{E} \in \mathbb{R}^{J \times R}$, $\mathbf{F} \in \mathbb{R}^{K \times R}$, as:
\begin{equation}\label{NN}
\mathbf{A}=\mathbf{D}*\mathbf{D}, \mathbf{B}=\mathbf{E}*\mathbf{E}, \mathbf{C}=\mathbf{F}*\mathbf{F}.
\end{equation}
Therefore, the problem \eqref{666666666} can be solved by optimization with regards to latent variables $\mathbf{D}$, $\mathbf{E}$, $\mathbf{F}$, which are linked to model factor matrices $\mathbf{A}$, $\mathbf{B}$, $\mathbf{C}$ by \eqref{NN}. Next, we will explain how to solve it by making the use of the NLS framework for C-CPD and the well-established chain rule for deriving derivatives.
\subsection{The NLS algorithm}
We use the NLS algorithm based on trust-region and Gauss-Newton (GN) methods which is of quadratic convergence when it approaches convergence to solve the problem \eqref{666666666}. The second-order Taylor expansion of objective function is:
\begin{equation*}\label{9}
{f}(\mathbf{x})\! \approx\! {f}({\mathbf{x}_k}) \!+\! \mathbf{p}_k^T {\frac{{\partial {f}({\mathbf{x}_k})}}{{\partial {\mathbf{x}_k}}}} \!+\! \frac{1}{2}\mathbf{p}_k^T{{\bf{H}}_k}\mathbf{p}_k^{},\left\| {\mathbf{p}_k^{}} \right\| \le {\Delta _k},
\vspace{-1mm}
\end{equation*}
with current iteration point $\mathbf{x} \buildrel \Delta \over= [\mathrm{vec}({\mathbf{D}});\mathrm{vec}({\mathbf{E}});\mathrm{vec}({\mathbf{F}})]$, step  ${\mathbf{p}_k} \buildrel \Delta \over= {\mathbf{x}} - {\mathbf{x}_k}$, Hessian $\mathbf{H}_k$ and trust-region radius $\Delta _k$.

Setting $\nabla_{\mathbf{p}_k} f(\mathbf{x}) = \mathbf{0}$ yields 
 $\mathbf{H}_{k} \mathbf{p}_{k} = -\mathbf{g}_k$  with $\mathbf{g}_k = {{\partial f(\mathbf{x}_k)} \mathord{\left/
 {\vphantom {{\partial f(\mathbf{x}_k)} {\partial \mathbf{x}_k}}} \right.
 \kern-\nulldelimiterspace} {\partial \mathbf{x}_k}}$.  Due to the large computation of the Moore-Penrose pseudo inverse of $\mathbf{H}_k$, we calculate a matrix-vector product (MVP) operation $\mathbf{H}_k\mathbf{z}_k$ in each preconditioned conjugated gradient (PCG) iteration. The preconditioned technique adopts a block diagonal matrix $\mathbf{N}_k$, which is selected to make $\mathbf{N}_k^{-1}\mathbf{H}_k$ close to $\mathbf{I}$, to reduce the number of iterations \cite{Pre}. The  steps of this algorithm are given in Algorithm 1. 
 In this table, ${\mathbf{D}}_{\rm{ini}}$, ${\mathbf{E}}_{\rm{ini}}$, ${\mathbf{F}}_{\rm{ini}}$ denote the initialized latent variables for ${\mathbf{D}}$, ${\mathbf{E}}$, ${\mathbf{F}}$. From the description of the NLS procedure, we know that two entities are key in to the algorithm, the gradient and Hessian of the $f$ with regards to the latent variables ${\mathbf{D}}$, ${\mathbf{E}}$, ${\mathbf{F}}$. In addition, the computation of MVP is also the key to PCG iterations.  So next we explain the computation of these two entities, as well as MVP.

\newlength{\ifindent}
\settowidth{\ifindent}{elll}
\begin{algorithm}[htb]
\caption{C-CPD-NLS-NN algorithm}\label{alg:}
\begin{algorithmic}[1]
\item[]\hspace{-1em}\textbf{Input: $\mathbf{P}_1$, $\mathbf{P}_2$, $\mathbf{P}_M$, ${\mathbf{D}}_{\rm{ini}}$, ${\mathbf{E}}_{\rm{ini}}$, ${\mathbf{F}}_{\rm{ini}}$, ${\boldsymbol{{\cal Y}}_H}$, ${\boldsymbol{{\cal Y}}_M}$}
{}
\item[]\hspace{-1em}\textbf{Output: $\mathbf{A}$, $\mathbf{B}$, $\mathbf{C}$}
\While{not converged}
    \State Compute gradient $\bf{g}$ by \eqref{13}-\eqref{18} \;
    \State Use PCG to solve $\mathbf{H} \mathbf{p} = -\mathbf{g}$ by matrix-vector product and preconditioner matrix to get the Newton point \;
    \State Solve the Cauchy point  by \cite{numerical} \;
    \State Update  trust-region radius $\Delta$ by \cite{numerical} \;
    \State Update $\mathbf{D}$, $\mathbf{E}$, $\mathbf{F}$ by $\Delta$, Cauchy point and Newton point
\EndWhile
\end{algorithmic}
\end{algorithm}

\subsection{Gradient and Jacobian}
Due to chain rule for deriving the derivatives, the gradient $\mathbf{g}$ can be calculated as follows:
\begin{equation}\label{13}
{\begin{aligned}
{\bf{g}} &={\frac{{\partial \mathbf{u}^T}}{{\partial {\mathbf{x}}}}}{\frac{{\partial f}}{{\partial {\mathbf{u}}}}}=[{\rm{vec}}({\nabla _{\bf{D}}}f);{\rm{vec}}({\nabla _{\bf{E}}}f);{\rm{vec}}({\nabla _{\bf{F}}}f)]\\
&= {\boldsymbol{\lambda} ^T}[{\rm{vec}}({\nabla _{\bf{A}}}f);{\rm{vec}}({\nabla _{\bf{B}}}f);{\rm{vec}}({\nabla _{\bf{C}}}f)],
\end{aligned}}
\end{equation}
where $\mathbf{u}=[\rm{vec}(\mathbf{A});\rm{vec}(\mathbf{B});\rm{vec}(\mathbf{C})]$ and $\boldsymbol{\lambda} = \rm{diag}(2\mathbf{u})$.
For convenience, we denote the factor matrices of $\boldsymbol{ \boldsymbol{ \boldsymbol{ \mathcal {Y}}}}_H$ by $[\kern-0.15em[ {\mathbf{U}^{(1)}},{\mathbf{U}^{(2)}},{\mathbf{U}^{(3)}}  ]\kern-0.15em]$ and those of $\boldsymbol{ \boldsymbol{ \mathcal {Y}}}_M$ by $[\kern-0.15em[ {\mathbf{V}^{(1)}},{\mathbf{V}^{(2)}},{\mathbf{V}^{(3)}} ]\kern-0.15em]$. Therefore, by \eqref{6}, we have ${\mathbf{U}^{(1)}}={\mathbf{P}_1}{\mathbf{A}}$, ${\mathbf{U}^{(2)}}={\mathbf{P}_2}{\mathbf{B}}$, ${\mathbf{U}^{(3)}}={\mathbf{C}}$, ${\mathbf{V}^{(1)}}={\mathbf{A}}$, ${\mathbf{V}^{(1)}}={\mathbf{B}}$ and ${\mathbf{V}^{(1)}}={\mathbf{P}_M}{\mathbf{C}}$. So $\nabla_\mathbf{A}f, \nabla_\mathbf{B} f, \nabla_\mathbf{C} f$ can be written as follows:
\begin{equation}\label{gradient}
    \begin{aligned}
    {\nabla _\mathbf{A}}{f} &= \mathbf{P}_1^T{\mathbf{G}_{H}^{(1)}} + {\mathbf{G}_{M}^{(1)}},\\
    {\nabla _\mathbf{B}}{f} &= \mathbf{P}_2^T{\mathbf{G}_{H}^{(2)}} + {\mathbf{G}_{M}^{(2)}},\\
    {\nabla _\mathbf{C}}{f} &= {\mathbf{G}_{H}^{(3)}} + \mathbf{P}_M^T{\mathbf{G}_{M}^{(3)}},
\end{aligned}
\end{equation}
where ${{\mathbf{G}}_{H}^{(n)}}$ and ${{\mathbf{G}}_{M}^{(n)}}$ can be defined in this way:
\begin{equation}\label{18}
    \begin{aligned}
    {{\mathbf{G}}_{H}^{(n)}} \buildrel \Delta \over= {{\bf{U}}^{(n)}}\mathbf{W}_{H}^{(n)T}{\mathbf{W}_{H}^{(n)}} - {\bf{Y}}_{H(n)}^{T}{\mathbf{W}_{H}^{(n)}},\\
    {{\mathbf{G}}_{M}^{(n)}} \buildrel \Delta \over= {{\bf{V}}^{(n)}}\mathbf{W}_{M}^{(n)T}{\mathbf{W}_{M}^{(n)}} - {\bf{Y}}_{M(n)}^{T}{\mathbf{W}_{H}^{(n)}},
\end{aligned}
\end{equation}
where ${\mathbf{W}_{H}^{(n)}}\buildrel \Delta \over=\odot _{q = 1,q \ne N - n + 1}^N{\mathbf{U}^{(N - q + 1)}}$, ${\mathbf{W}_{M}^{(n)}}$ is analogous to that of ${\mathbf{W}_{H}^{(n)}}$.

For GN, the Hessian $\mathbf{H}$ is approximated with the Gramian of the Jacobian i.e. $\mathbf{H} = \mathbf{H}_H+\mathbf{H}_M={\bf{J}}_H^T{{\bf{J}}_H} + {\bf{J}}_M^T{{\bf{J}}_M}$, where $\mathbf{J}_H$ and $\mathbf{J}_M$ denote the Jacobian matrix. 
\begin{equation}\label{Jaci}
    \begin{aligned}
    {{\bf{J}}_H}&=(\partial {\rm{vec}}({\boldsymbol{{\cal Y}}_H}-[\kern-0.15em[ {{\bf{P}}_1}{\bf{A}},{{\bf{P}}_2}{\bf{B}},{\bf{C}} ]\kern-0.15em]_R )/\partial {{\bf{x}}^T}),\\
    {{\bf{J}}_M}& = (\partial {\rm{vec}}({\boldsymbol{{\cal Y}}_M} -  [\kern-0.15em[ {\bf{A}},{\bf{B}},{{\bf{P}}_M}{\bf{C}} ]\kern-0.15em]_R )/\partial {{\bf{x}}^T}),
\end{aligned}
\end{equation}
%
\subsection{Matrix-vector product}

Then, we derive an efficient method to compute the product of $\mathbf{H}$ and a vector $\mathbf{z}$. Based on  
 the chain rule for deriving the derivatives,
\begin{equation}\label{20}
\mathbf{H}\mathbf{z} = \boldsymbol{\lambda}^T(\mathbf{K}^T{\mathbf{K}} + \mathbf{M}^T{\mathbf{M}})\boldsymbol{\lambda}\mathbf{z},
\end{equation}
where $\mathbf{K}\buildrel \Delta \over= [\mathbf{K}_1,\mathbf{K}_2,\mathbf{K}_3]$, $\mathbf{M}\buildrel \Delta \over= [\mathbf{M}_1,\mathbf{M}_2,\mathbf{M}_3]$ also denote the Jacobian matrices \eqref{Jaci2} and the vector $\boldsymbol{\lambda}\mathbf{z}$ can be written as $\mathbf{t}=[\text{vec}({\mathbf{T}_{1}});\text{vec}({\mathbf{T}_{2}});\text{vec}({\mathbf{T}_{3}})]\in {\mathbb{R}^{(I+J+K)R\times 1} }$. 
\begin{equation}\label{Jaci2}
    \begin{aligned}
    &{\bf{K}}=(\partial {\rm{vec}}({\boldsymbol{{\cal Y}}_H}-[\kern-0.15em[ {{\bf{P}}_1}{\bf{A}},{{\bf{P}}_2}{\bf{B}},{\bf{C}} ]\kern-0.15em]_R )/\partial {{\bf{u}}^T}),\\
    &{\bf{M}} = (\partial {\rm{vec}}({\boldsymbol{{\cal Y}}_M} -  [\kern-0.15em[ {\bf{A}},{\bf{B}},{{\bf{P}}_M}{\bf{C}} ]\kern-0.15em]_R )/\partial {{\bf{u}}^T}).
\end{aligned}
\end{equation}
We take the calculation of 
$\mathbf{K}^T{\mathbf{K}}\mathbf{t}$ as an example. 
Hence,
\begin{equation}
    \mathbf{K}_{{n_1}}^T\mathbf{K}_{{n_2}}\mathbf{t}\buildrel \Delta \over = \mathrm{vec} \left(\mathbf{R}_{n_1}\mathbf{L}_{(n_1,n_2)}\right),
\end{equation}
$\mathbf{R}_{n_1}=\mathbf{P}_{n_1}$ when $n_1=1,2$, $\mathbf{R}_{n_1}=\mathbf{I}_K$ when $n_1 = 3$.
\begin{equation*}\label{25}
\mathbf{L}_{(n_1,n_2)}\!=\!\left\{\! 
{\begin{aligned}
&{\mathbf{T}_{n}}\mathbf{W}_{H}^{(n)T}\mathbf{W}_{H}^{(n)}, {n_1} = {n_2} = n = 1,2,3,\\
&\mathbf{U}^{({n_1})}(\mathbf{T}_{{n_2}}^T\mathbf{U}^{({n_2})}*\mathbf{U}^{(n)T}\mathbf{U}^{(n)}),\\
& \;\;\;\;\;\;\;\;\;\;\;\;\;\;\;\;n_1\neq n_2, n = \{1,2,3\}\backslash   \{n_1,n_2\}. \\
\end{aligned}} \right.
\end{equation*}
$\mathbf{M}^T{\mathbf{M}}\mathbf{t}$ is analogous to that of $\mathbf{K}^T{\mathbf{K}}\mathbf{t}$.
\section{EXPERIMENT RESULTS}
\label{sec:typestyle}
We compare the proposed NN-C-CPD-NLS algorithm with unconstrained C-CPD-ALS (UC-C-CPD-ALS) \cite{STEREO}, hyperspectral super-resolution via subspace-based regularization (HYSURE) \cite{hysure},  fast fusion based on sylvester equation (FUSE) \cite{fuse}, UC-C-CPD-NLS, NN-C-BTD-APG \cite{C-BTD} and UC-C-Tucker \cite{C-Tucker} algorithms. In this paper, we use the root mean square error (RMSE), reconstruction signal-to-noise ratio (R-SNR), cross correlation (CC) and spectral angle mapper (SAM) to evaluate the quality of the recovered SRI \cite{metric}.
\[\text{RMSE} ={\parallel {{\hat {\boldsymbol{ \mathcal {Y}}}}_S}{ - }{{\boldsymbol{ \mathcal {Y}}}_S}\parallel _F}/{{\sqrt {IJK} }},\]
\[\text{CC} = {{1}/{K}}\sum\nolimits_{k{ = 1}}^K {\rho ({{\hat {\boldsymbol{ \mathcal {Y}}}}_S}{(:,:,}k{),}{{\boldsymbol{ \mathcal {Y}}}_S}{(:,:,}k{)})} ,\]
where $\rho$ is the Pearson correlation coefficient between  ${\hat {\boldsymbol{ \mathcal {Y}}}_S}$ and  ${{\boldsymbol{ \mathcal {Y}}}_S}$. CC is a score between $0$ and $1$, and $1$ is the best result.
\[\text{R-SNR=10log}_{10}\left( {\frac{{\sum\limits_{k = 1}^K {\parallel {{\boldsymbol{ \mathcal {Y}}}_S}{(:,:,}k{)}\parallel _F^{2}} }}{{\sum\limits_{k = 1}^K {\parallel {{\hat {\boldsymbol{ \mathcal {Y}}}}_S}{(:,:,}k{) - }{{\boldsymbol{ \mathcal {Y}}}_S}{(:,:,}k{)}\parallel _F^{2}} }}} \right).\]
The higher the R-SNR, the better the SRI recovery.
\[
\text{SAM} = \frac{1}{IJ} \sum_{n=1}^{IJ} \arccos \left( \frac{\mathbf{Y}_S^{(3)}(n,:) \cdot \hat{\mathbf{Y}}_S^{(3)}(n,:)^T}{\|\mathbf{Y}_S^{(3)}(n,:)\| \|\hat{\mathbf{Y}}_S^{(3)}(n,:)\|} \right).
\]
SAM measures the angle between estimated and ground-truth SRI fibers, with smaller values indicating better performance.

\begin{figure}[h]
    \centering
        \includegraphics[width=0.95\columnwidth]{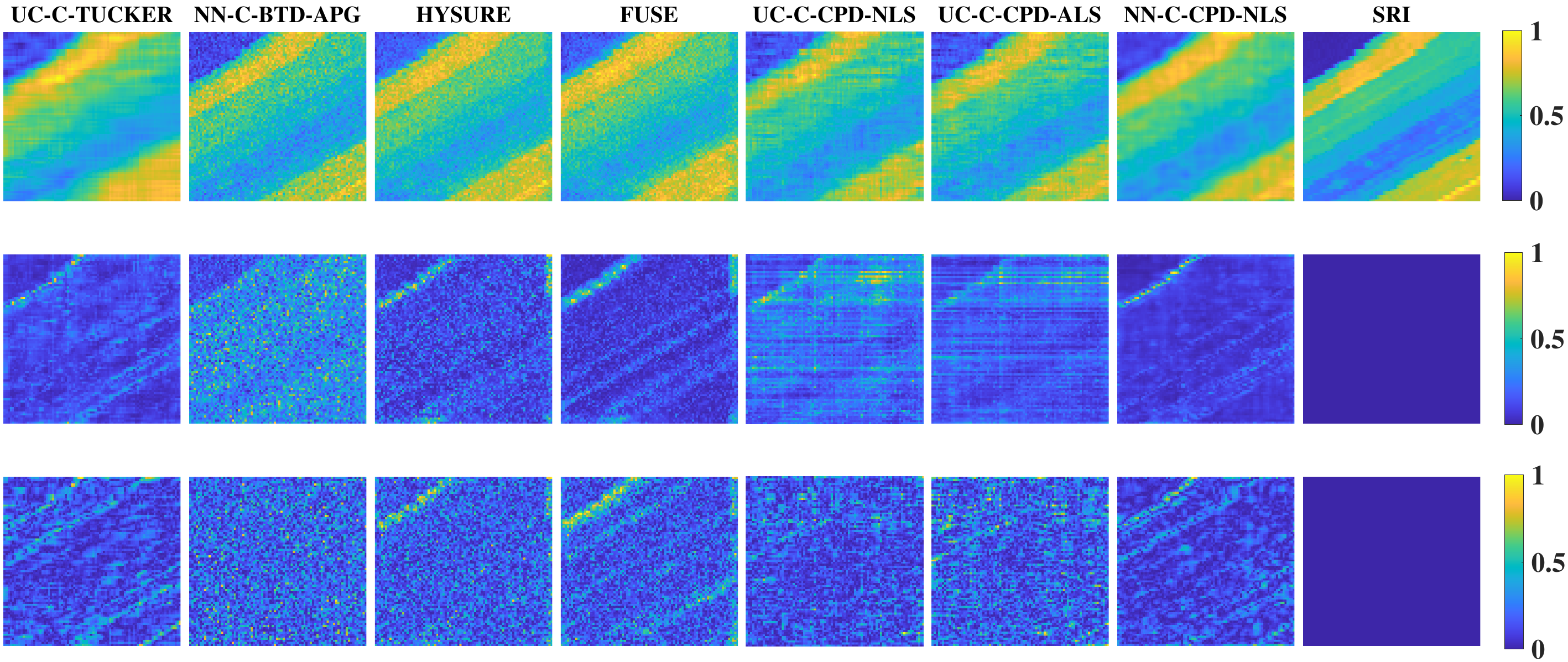}
    \captionsetup{aboveskip=0pt, belowskip=0pt} 
    \vspace{3mm} 
\caption{First row: the recovered SRIs of the $34$-th band; Second row: the residual images of the $34$-th band; Third row: the SAM maps}
    \label{fig:RESULT}
    \vspace{-3mm}
\end{figure}

In the first experiment, we use the dataset from the airborne visible/infrared imaging spectrometer (AVIRIS) platform as SRI i.e. ${{\boldsymbol{ \mathcal {Y}}}_S} \in \mathbb{R}^{80 \times 84 \times 204}_{+}$. This scene depicts a field containing six different agricultural products, measured across 204 spectral bands. We compare several algorithms using this data with an SNR of $5 \text{dB}$.  The HSI is modeled by a $9 \times 9$ Gaussian kernel, which plays a role in spatial blurring and the blurred image is downsampled by a factor of $d = 4$ in both spatial dimensions \cite{STEREO} i.e. ${{\boldsymbol{ \mathcal {Y}}}_H} \in \mathbb{R}^{20 \times 21 \times 204}_{+}$. MSI is generated by the band selection and averaging operation detailed in \cite{STEREO} i.e. ${{\boldsymbol{ \mathcal {Y}}}_M} \in \mathbb{R}^{80 \times 84 \times 6}_{+}$. The NN-C-CPD-NLS, UC-C-CPD-NLS and UC-C-CPD-ALS algorithms use a rank of $30$, while the matrix based and NN-C-BTD-APG methods use a rank of $6$ which is the same as \cite{STEREO,C-BTD}. For UC-C-Tucker algorithm, we use $[12, 12, 12]$ as the multilinear rank. The initialization method of the UC-C-CPD-ALS algorithm is the same as in \cite{STEREO}, while other tensor-based algorithms are initialized randomly and non-negatively. Fig. \ref{fig:RESULT} shows the $34$th band of the estimated SRI, the residual images, and the SAM maps \cite{metric}. To enhance performance, we applied spatial smoothing to each slice in the spatial dimension of the reconstructed SRI \cite{smooth} which is also used in other experiments. Clearly, the proposed algorithm perform better at low SNR.

\begin{figure}[h]
    \vspace{-3pt}
    \centering
    \begin{minipage}{0.48\columnwidth}
        \centering
        \includegraphics[width=\textwidth]{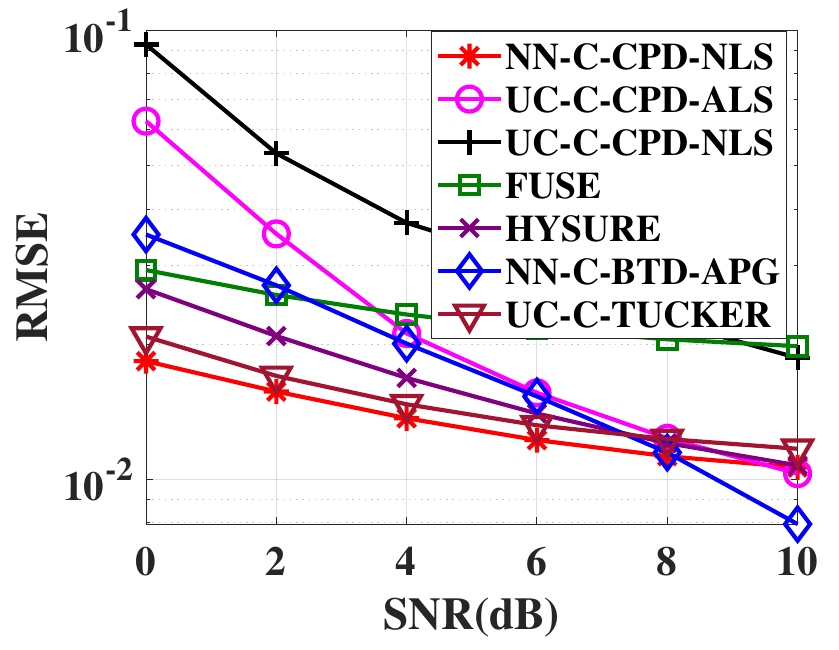}
        \centerline{(a) RMSE}
    \end{minipage}
    \begin{minipage}{0.48\columnwidth}
        \centering
        \includegraphics[width=\textwidth]{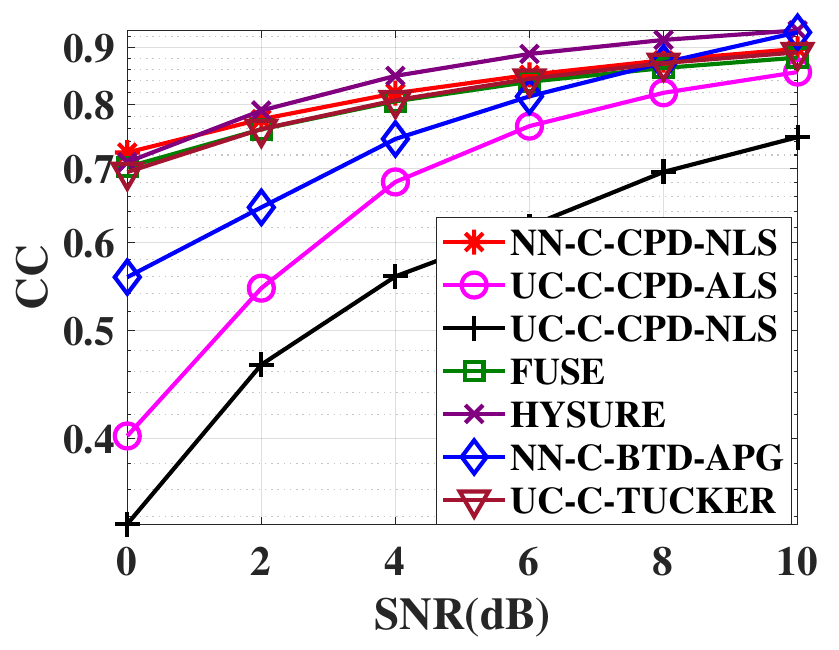}
        \centerline{(b) CC}
        \vspace{-3mm}
    \end{minipage}
       \begin{minipage}{0.48\columnwidth}
        \centering
        \includegraphics[width=\textwidth]{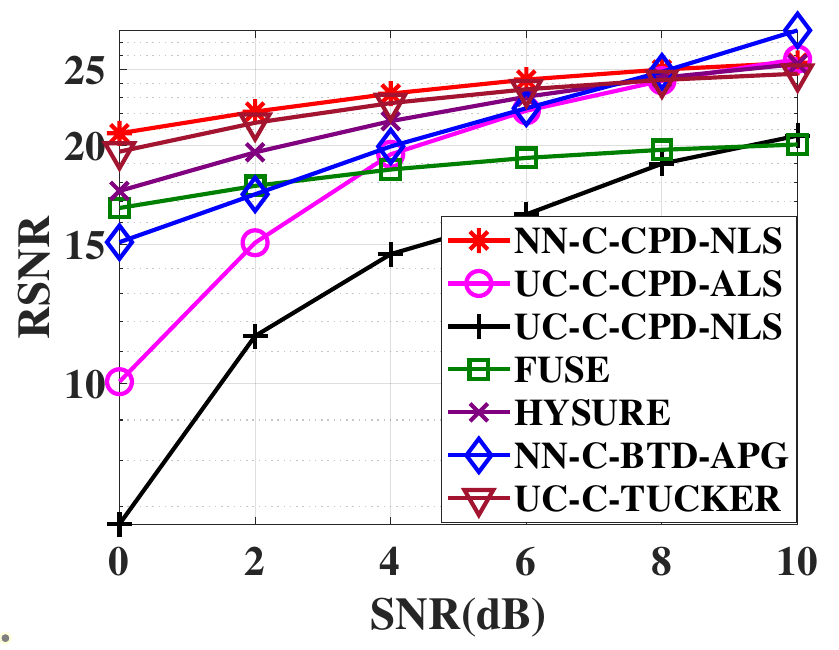}
        \centerline{(c) R-SNR}
    \end{minipage}
        \begin{minipage}{0.48\columnwidth}
        \centering
        \includegraphics[width=\textwidth]{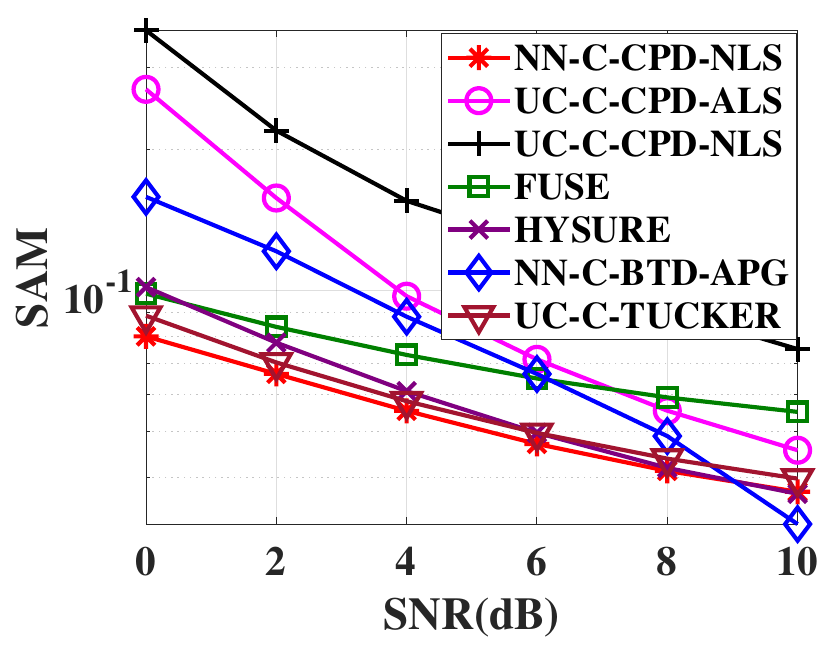}
        \centerline{(d) SAM}
    \end{minipage}
    \captionsetup{aboveskip=0pt, belowskip=0pt}
    \vspace{3mm}
    \caption{Reconstruction metrics for SRI under different SNR}
    \label{fig:SNR}
\end{figure}

In the second experiment, we use the same data as before. We vary the SNR from $0\text{dB}$ to $10\text{dB}$ and conduct a Monte Carlo experiment consisting of 50 runs in each SNR. The generation of HSI and MSI, methods of  initialization and rank of these algorithms are the same as before. Metrics are calculated as the median of the results from 50 Monte Carlo runs. Fig. \ref{fig:SNR} shows our algorithm has better performance in noisy situations compared to other algorithms.

\begin{figure}[h]
    \vspace{-3pt}
    \centering
    \begin{minipage}{0.48\columnwidth}
        \centering
        \includegraphics[width=\textwidth]{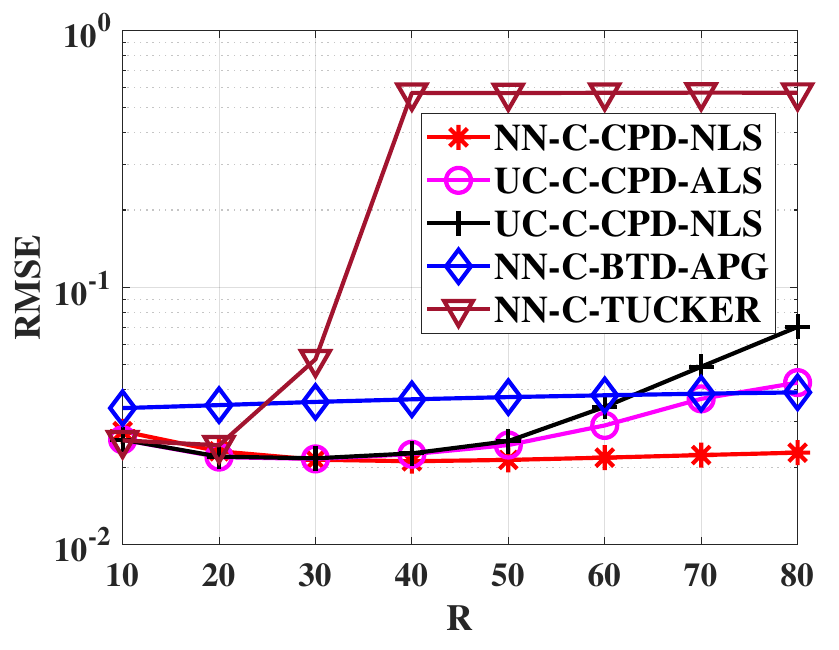}
        \centerline{(a) RMSE}
    \end{minipage}
    \begin{minipage}{0.48\columnwidth}
        \centering
        \includegraphics[width=\textwidth]{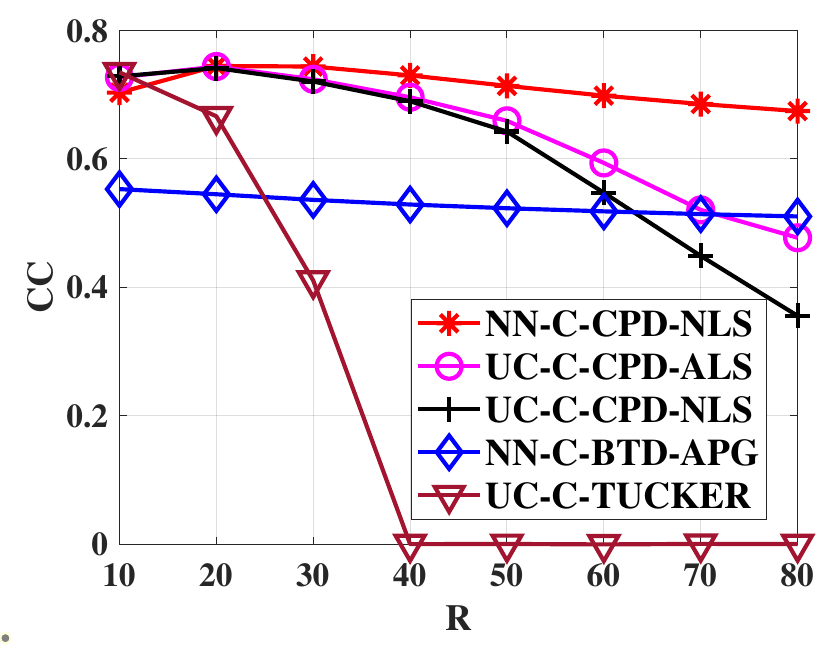}
        \centerline{(b) CC}
        \vspace{-3mm}
    \end{minipage}
       \begin{minipage}{0.48\columnwidth}
        \centering
        \includegraphics[width=\textwidth]{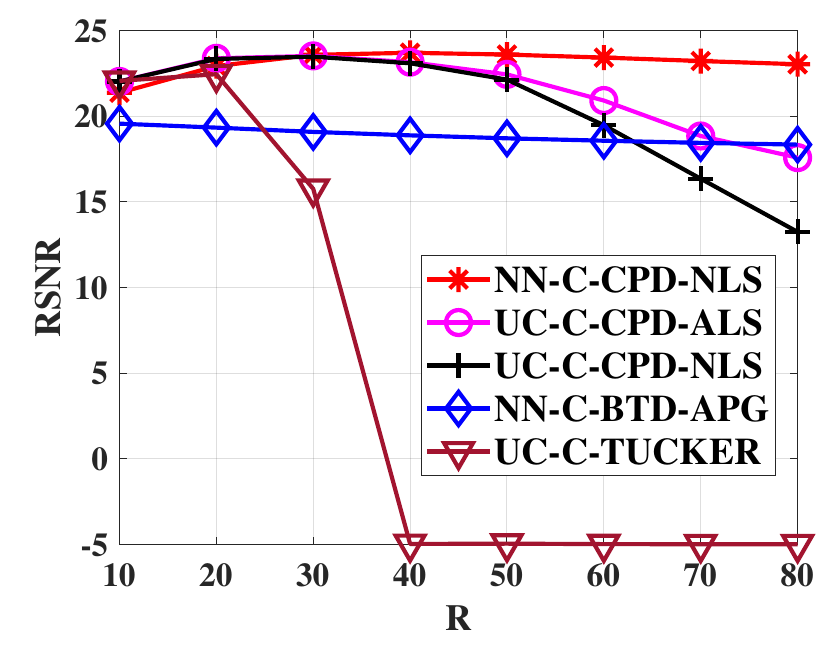}
        \centerline{(c) R-SNR}
    \end{minipage}
        \begin{minipage}{0.48\columnwidth}
        \centering
        \includegraphics[width=\textwidth]{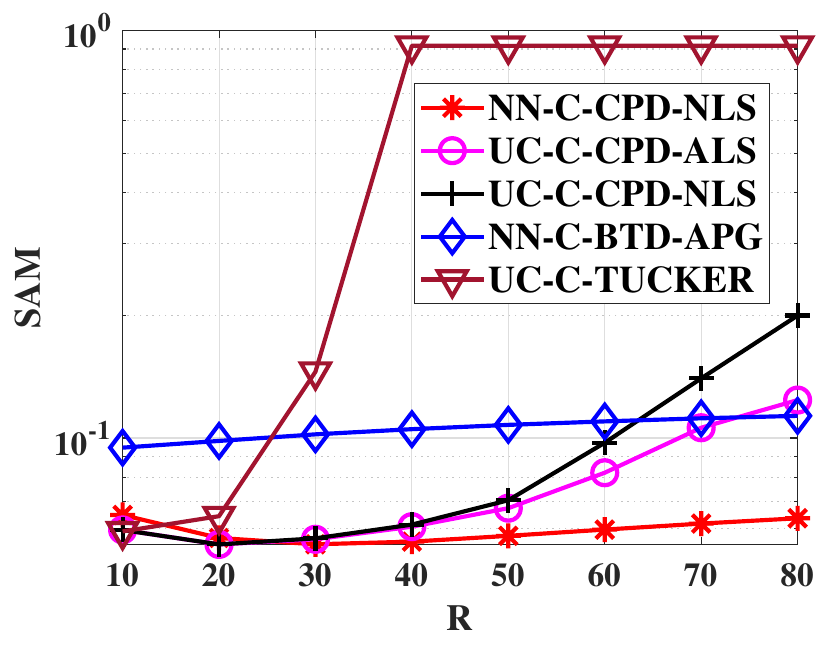}
        \centerline{(d) SAM}
    \end{minipage}
    \captionsetup{aboveskip=0pt, belowskip=0pt}
    \vspace{3mm}
    \caption{Reconstruction metrics for SRI under different $R$}
    \label{fig:R}
\end{figure}

In the third experiment, we use another data  with agriculture, forest, and natural vegetation from the AVIRIS platform, namely, the Indian Pines as the SRI. The HSI and MSI are generated as before. We vary the multilinear rank of UC-C-Tucker algorithm from $[10,10,10]$ to $[80,80,80]$. We vary the $R$ for  other tensor decomposition algorithms from 10 to 80. 
With a fixed SNR of $5\text{dB}$ and the same initialization as before, we compare five tensor-based algorithms. Fig. \ref{fig:R} shows our algorithm maintains stable performance across different $R$ values due to the parametric method to add NN constrains.

\vspace{-1mm}

\vspace{-3mm}
\vspace{-6mm}
\section{CONCLUSION}
\label{sec:majhead}

We propose a parametric non-negative (NN) coupled CPD algorithm based on NLS framework for HSR. The proposed algorithm converts the NN constraint into the squared relationship between the
NN entries of the factor matrices and a set of latent parameters. Therefore, the NN constraint is naturally integrated without interfering with the optimization
procedure in the proposed algorithm. Experiment results show that the proposed algorithm can better recover SRI in high noise for different $R$, compared with several existing coupled tensor/matrix based algorithms.




\bibliographystyle{IEEEbib}
\bibliography{strings,refs}

\begin{thebibliography}{10}

\bibitem{STEREO}
Charilaos~I. Kanatsoulis, Xiao Fu, Nicholas~D. Sidiropoulos, and Wing-Kin Ma,
\newblock ``{Hyperspectral Super-Resolution}: A coupled tensor factorization
  approach,''
\newblock {\em IEEE Trans. Signal Process.}, vol. 66, no. 24, pp. 6503--6517,
  Dec. 2018.

\bibitem{nonCCPD}
Yang Xu, Zebin Wu, Jocelyn Chanussot, Pierre Comon, and Zhihui Wei,
\newblock ``Nonlocal coupled tensor \text{CP} decomposition for hyperspectral
  and multispectral image fusion,''
\newblock {\em IEEE Trans. Geosci. Remote Sens.}, vol. 58, pp. 348--362, Jan.
  2020.

\bibitem{C-Tucker}
Pr{\'e}vost Cl{\'e}mence, Usevich Konstantin, Pierre Comon, and David Brie,
\newblock ``Hyperspectral super-resolution with coupled \text{Tucker}
  approximation: Recoverability and svd-based algorithms,''
\newblock {\em IEEE Trans. Signal Process.}, vol. 68, pp. 931--946, Jan. 2020.

\bibitem{TD-TUCKER-NN}
Wei Wan, Weihong Guo, Haiyang Huang, and Jun Liu,
\newblock ``Nonnegative and nonlocal sparse tensor factorization-based
  hyperspectral image super-resolution,''
\newblock {\em IEEE Trans. Geosci. Remote Sens.}, vol. 58, no. 12, pp.
  8384--8394, Apr. 2020.

\bibitem{C-BTD}
Meng Ding, Xiao Fu, Ting-Zhu Huang, Jun Wang, and Xi-Le Zhao,
\newblock ``Hyperspectral super-resolution via interpretable block-term tensor
  modeling,''
\newblock {\em IEEE J. Sel. Top. Signal Process.}, vol. 15, no. 3, pp.
  641--656, Apr. 2021.

\bibitem{CBTDNN2024}
Clémence Prévost and Valentin Leplat,
\newblock ``Minimum-volume non-negative block-term decomposition: Blind data
  fusion and unmixing with estimation of the number of endmembers,''
\newblock in {\em 2024 32nd European Signal Processing Conference (EUSIPCO)},
  Oct. 2024, pp. 1057--1061.

\bibitem{review}
Naoto Yokoya, Claas Grohnfeldt, and Jocelyn Chanussot,
\newblock ``Hyperspectral and multispectral data fusion: A comparative review
  of the recent literature,''
\newblock {\em IEEE Geosci. Remote Sens. Mag.}, vol. 5, no. 2, pp. 29--56, Jun.
  2017.

\bibitem{ADMM}
Athanasios~P. Liavas and Nicholas~D. Sidiropoulos,
\newblock ``Parallel algorithms for constrained tensor factorization via
  alternating direction method of multipliers,''
\newblock {\em IEEE Trans. Signal Process.}, vol. 63, no. 20, pp. 5450--5463,
  Jul. 2015.

\bibitem{FCCPD}
Rodrigo~Cabral Farias, Jeremy~E. Cohen, and Pierre Comon,
\newblock ``Exploring multimodal data fusion through joint decompositions with
  flexible couplings,''
\newblock {\em IEEE Trans. Signal Process.}, vol. 64, pp. 4830--4844, Sept.
  2015.

\bibitem{NNMATRIX}
Weisheng Dong, Fazuo Fu, Guangming Shi, Xun Cao, Jinjian Wu, Guangyu Li, and
  Xin Li,
\newblock ``Hyperspectral image super-resolution via non-negative structured
  sparse representation,''
\newblock {\em IEEE Transactions on Image Processing}, vol. 25, no. 5, pp.
  2337--2352, Mar. 2016.

\bibitem{NNRS}
Tatsuya Yokota, Rafal Zdunek, Andrzej Cichocki, and Yukihiko Yamashita,
\newblock ``Smooth nonnegative matrix and tensor factorizations for robust
  multi-way data analysis,''
\newblock {\em Signal Process.}, vol. 113, pp. 234--249, Aug. 2015.

\bibitem{NicNNNLS}
N.~Vervliet,
\newblock {\em Compressed Sensing Approaches to Large-Scale Tensor
  Decompositions},
\newblock Ph.D. thesis, KU Leuven, 5 2018.

\bibitem{NLS}
Jean-Philip Royer, Nad{\`e}ge Thirion-Moreau, and Pierre Comon,
\newblock ``Computing the polyadic decomposition of nonnegative third order
  tensors,''
\newblock {\em Signal Process.}, vol. 91, pp. 2159--2171, Sept. 2011.

\bibitem{compress}
Miguel~A Veganzones, Jeremy~E Cohen, Rodrigo~Cabral Farias, Jocelyn Chanussot,
  and Pierre Comon,
\newblock ``Nonnegative tensor \text{CP} decomposition of hyperspectral data,''
\newblock {\em IEEE Trans. Geosci. Remote Sens.}, vol. 54, no. 5, pp.
  2577--2588, Dec. 2015.

\bibitem{Pre}
Martijn Bouss{\'e}, Nico Vervliet, Ignat Domanov, Otto Debals, and Lieven~De
  Lathauwer,
\newblock ``Linear systems with a canonical polyadic decomposition constrained
  solution: Algorithms and applications,''
\newblock {\em Numer. Linear Algebra Appl.}, vol. 25, Aug. 2018.

\bibitem{numerical}
J~Nocedal and SJ~Wright,
\newblock {\em Numerical Optimization},
\newblock Springer, New York, 2nd edition, 2006.

\bibitem{hysure}
Miguel Simões, José Bioucas‐Dias, Luis~B. Almeida, and Jocelyn Chanussot,
\newblock ``A convex formulation for hyperspectral image superresolution via
  subspace-based regularization,''
\newblock {\em IEEE Trans. Geosci. Remote Sens.}, vol. 53, no. 6, pp.
  3373--3388, Jun. 2015.

\bibitem{fuse}
Qi~Wei, Nicolas Dobigeon, and Jean-Yves Tourneret,
\newblock ``Fast fusion of multi-band images based on solving a sylvester
  equation,''
\newblock {\em IEEE Trans. Image Process.}, vol. 24, no. 11, pp. 4109--4121,
  Nov. 2015.

\bibitem{metric}
Laetitia Loncan, Luis~B. de~Almeida, Jose~M. Bioucas-Dias, Xavier Briottet,
  Jocelyn Chanussot, Nicolas Dobigeon, Sophie Fabre, Wenzhi Liao, Giorgio~A.
  Licciardi, Miguel Simões, Jean-Yves Tourneret, Miguel~Angel Veganzones,
  Gemine Vivone, Qi~Wei, and Naoto Yokoya,
\newblock ``Hyperspectral pansharpening: A review,''
\newblock {\em IEEE Geosci. Remote Sens. Mag.}, vol. 3, no. 3, pp. 27--46,
  Sept. 2015.

\bibitem{smooth}
Jong-Sen Lee,
\newblock ``Digital image enhancement and noise filtering by use of local
  statistics,''
\newblock {\em IEEE Trans. Pattern Anal. Mach. Intell.}, vol. PAMI-2, no. 2,
  pp. 165--168, Mar. 1980.

\end{thebibliography}

\end{document}